\documentstyle[prl,aps,epsfig]{revtex}
\begin{document}
\twocolumn[\hsize\textwidth\columnwidth\hsize\csname@twocolumnfalse\endcsname

\date{}
\title{Magnetism and superconductivity in CeRh$_{1-x}$Ir$_x$In$_5$
     heavy fermion materials.}
\author{G.D.~Morris${^1}$, R.H.~Heffner${^1}$, J.E.~Sonier${^2}$, D.E.~MacLaughlin${^3}$,
O.O.~Bernal${^4}$, G.J.~Nieuwenhuys${^5}$, A.T.~Savici${^6}$, P.G.~Pagliuso${^1}$ and J.L.~Sarrao${^1}$}
\address{${^1}$Los Alamos National Laboratory,MS K764,Los Alamos, NM 87545,USA\\
${^2}$Department of Physics, Simon Fraser University, Burnaby, British Columbia V5A 1S6, Canada\\
${^3}$Department of Physics, University of California, Riverside, California 92521, USA\\
${^4}$Physics Department, California State University, Los Angeles, California 90032, USA\\
${^5}$Kamerlingh Onnes Laboratory, Leiden University,  Postbus 9504 - 2300 RA Leiden, The Netherlands\\
${^6}$Department of Physics, Columbia University,  New York, NY 10027, USA}
\date{\today}
\maketitle
\begin{abstract}
We report zero-field $\mu$SR studies of cerium based heavy-fermion
materials CeRh$_{1-x}$Ir$_x$In$_5$. 
In the superconducting $x$=0.75 and 1 compositions muon spin 
relaxation functions were found to be temperature independent 
across $T_{\rm c}$;
no evidence for the presence of electronic magnetic moments
was observed.
The $x$=0.5 material is antiferromagnetic below $T_{\rm N}$=3.75~K 
and superconducting below  $T_{\rm c}$=0.8~K.  $\mu$SR
spectra show the gradual onset of damped coherent
oscillations characteristic of magnetic order below $T_{\rm N}$.
At 1.65~K the total oscillating amplitude accounts for at least 85\% of
the sample volume. 
No change in precession frequency or amplitude is
detected on cooling below $T_{\rm c}$, indicating the microscopic
coexistence of magnetism and superconductivity in this material.
\end{abstract}
\pacs{PACS numbers: }
]

A tetragonal HoCoGa$_5$ crystal structure is common to all the
CeMIn$_5$ materials, which may be viewed as alternating layers
of CeIn$_3$ and MIn$_2$. 
Varying fractions of Ir, Rh or Co on the M site
results in a rich phase diagram exhibiting 
antiferromagnetic and superconducting properties.
It has been noted \cite{Pagliuso-2001} that the superconducting 
transition temperature $T_{\rm c}$ increases
monotonically with $c/a$,  suggesting that
the superconductivity is most robust when the layers of
CeIn$_3$ are more isolated. It is not clear whether this
reflects a shift toward optimal hybridization, or
greater 2D character in a possible magnetic pairing interaction.
CeCoIn$_5$ has the highest $T_{\rm c}$ = 2.3~K of the 
known heavy fermion superconductors \cite{Petrovic-2001}.
Specific heat and thermal conductivity measurements
on CeCoIn$_5$ and CeIrIn$_5$ found power-law temperature
dependences, indicating the presence of lines of nodes
in their superconducting gaps \cite{Movshovich-2001}.
These results led to speculation that 2D (or at least anisotropic)
AFM fluctuations play an important role in magnetically-mediated
heavy-fermion superconductivity within the CeMIn$_5$ family.
Recent theoretical work has also found that quasi-2D magnetic
fluctuations result in superconductivity at higher temperatures
than 3D fluctuations in metals which are nearly
antiferromagnetic \cite{Monthoux-2001}.
However, neutron scattering experiments on CeRhIn$_5$ found
AFM fluctuations along the $c$-axis, with a weaker
interaction than between Ce moments in-plane,
but concluded that magnetic correlations are nonetheless 3D 
in nature \cite{Bao-2002}.
We have performed $\mu$SR experiments on several
superconducting compositions of CeRh$_{1-x}$Ir$_x$In$_5$, with
the objective of elucidating the magnetic and superconducting
properties of these materials.

The phase diagram of CeRh$_{1-x}$Ir$_x$In$_5$ shows a range
of magnetic and superconducting properties which may be tuned
with composition or pressure \cite{Hegger-2000}.
CeRhIn$_5$ is known from neutron scattering experiments
to have an incommensurate antiferromagnetic
order below $T_{\rm N}$=3.8~K, in which the moments lie
in the $a-b$ plane but spiral 107$^{\circ}$ per unit cell
along the $c$-axis \cite{Bao-AFM-2000}.
Superconductivity is present for compositions 0.3 $<x<$ 1,
and there is evidence from macroscopic susceptibility and 
resistivity measurements that AFM order and
superconductivity coexist for 0.3 $<x<$ 0.6 \cite{Pagliuso-2001}.
A local probe like $\mu$SR can address whether
the magnetism and superconductivity coexist on a
microscopic scale.

Zero-field time-differential $\mu$SR spectra were taken on superconducting
samples of CeIrIn$_5$ ($T_{\rm c}$=0.4~K) and CeRh$_{0.25}$Ir$_{0.75}$In$_5$
($T_{\rm c}$=0.7~K) with the dilution refrigerator at the {\sc triumf} M15 
channel.
Zero magnetic field (ZF) conditions were prepared by quenching the
superconducting solenoid in the dilution refrigerator. 
Residual fields on the order of 1~G were then zeroed to less than
20~mG with external trim coils and a field-zeroing
method described elsewhere in these proceedings \cite{Mu-FZ}.
Samples were oriented with their $c$ axes nominally parallel
to the muon momentum and spin.

A background signal originating from muons stopping
in cryostat parts near the sample position was
characterized with the entire area of the silver sample 
stage covered by a GaAs wafer, which produces no decay
asymmetry in ZF.
In both CeIrIn$_5$ and CeRh$_{0.25}$Ir$_{0.75}$In$_5$ the
experimental decay asymmetry followed a slow, static Gaussian
Kubo-Toyabe form with temperature-independent relaxation rates
$\Delta$ across their superconducting transitions, shown in Fig.~1.
Average values of $\Delta$ in CeIrIn$_5$ and
CeRh$_{0.25}$Ir$_{0.75}$In$_5$ were found to be
0.239(4) and 0.276(2) $\mu$s$^{-1}$ respectively.
There is thus no evidence of an increased relaxation rate
or coherent precession signal which might result
from the onset of electronic (spin or orbital) moments
at $T_{\rm c}$.
These results are similar to those recently published by Higemoto 
{\em et.\/al.}\cite{Higemoto-2002} who also found no observable
electronic magnetic moment in the superconducting state in
CeIrIn$_5$.
Our results differ in that we obtained excellent fits of
the polarization to a {\em static} Kubo-Toyabe relaxation
function.

Zero-field experiments were performed on 
two samples of CeRh$_{0.5}$Ir$_{0.5}$In$_5$ 
at the LTF ($T\leq$1.15~K) and GPS ($T\geq$1.65~K) channels 
at PSI, Switzerland. 
Auxiliary experiments found no indications of inhomogeneity
in these samples;  sharp transitions were found in bulk 
resistivity and AC susceptibility.
$\mu$SR spectra (Fig.~2a) show rapidly-damped, coherent oscillations 
developing below $T_{\rm N}$=3.75~K. 
This component of the signal is clearly not of a
Kubo-Toyabe form since the minimum in corrected 
asymmetry is negative.
From $T_{\rm N}$ down to 2~K the static Gaussian Kubo-Toyabe
(G-KT) component diminishes in amplitude, while the oscillating
and longitudinal components grow.
Above $T$=2~K the experimental asymmetry was fitted to the
sum of two oscillating signals, plus the slow G-KT
and a longitudinal signal due to that component of the muon
spin parallel to the local magnetic field.
The total asymmetry remained constant.
The G-KT signal vanishes at temperatures of 1.65~K and below.  
For spectra taken in the dilution refrigerator, a non-relaxing 
silver background component was also present. Figure 2b shows
examples of two spectra at 0.1 and 1.15 K.

Two oscillating components were observed in the first
$\mu$s of the CeRh$_{0.5}$Ir$_{0.5}$In$_5$ spectra taken at temperatures 
below 2.75~K.
Total asymmetry of the oscillating components saturated by $T$=2~K.
Rapid damping made it impossible to obtain reliable fits to the higher
frequency and its relaxation rate, however.
The muon site(s) in these materials remain to be determined.
Nevertheless, transverse field data in CeIrIn$_5$ from 
$T$=2--300~K \cite{TF-Ce115} show that the muon stops in
two magnetically inequivalent sites.
This is consistent with the two observed oscillating signals in
CeRh$_{0.5}$Ir$_{0.5}$In$_5$.
High relaxation rates may be the result of a broad field
distribution at the muon sites which would result from 
incommensurate magnetic order similar to that in CeRhIn$_5$.

The smaller frequency in CeRh$_{0.5}$Ir$_{0.5}$In$_5$
($\nu_1(T)$ in Fig.~3) is nearly temperature-independent below 
1.15~K, with no obvious discontinuity at $T_{\rm c}$.
The oscillating asymmetry in spectra taken with the muon spin
rotated (to 50$^{\circ}$ from the $c$-axis)
amounted to 85\% of the full asymmetry so {\em at least} 
85\% of the sample volume is magnetic.
Spectra were essentially indistinguishable between $T$=1.15~K and
0.050~K, but these spectra have smaller oscillating amplitudes and
larger longitudinal amplitudes due to the orientation of the muon
spin with respect to the local field.
We have therefore normalized the total oscillating
fraction measured in spectra for $T\leq$1.15~K to those
measured with rotated spin at $T\geq$1.65~K.
This corrected oscillating fraction is also shown in Fig.~3.
Note that no change occurred in the asymmetry or frequency
on crossing the superconducting transition - {\it i.e.,} magnetic
order persists unchanged into the superconducting state.
Susceptibility measurements reveal a diamagnetic signal at
$T_{\rm c}$ that is at least 85\% of that expected for full shielding.
We, therefore, conclude that superconductivity and AFM order 
must coexist microscopically in the majority of the sample
volume.

In summary, we find no evidence for electronic magnetic moments
coincident with the onset of superconductivity in 
CeIrIn$_5$ or CeIr$_{0.75}$Rh$_{0.25}$In$_5$ 
Furthermore, the antiferromagnetism in CeIr$_{0.5}$Rh$_{0.5}$In$_5$ 
coexists microscopically with superconductivity below T$_{\rm c}$.

This work was supported in part by US NSF grants 
DMR-0102293 (UC Riverside) and DMR-9820631 (CSU Los Angeles).
JES acknowledges support from the Natural Sciences and Engineering
Research Council (NSERC) and the Canadian Institute for Advanced
Research. Work at Los Alamos was performed under the auspices
of the U.S. Department of Energy, Office of Science.

\section*{Figure captions}
Fig.1 ZF static Gaussian Kubo-Toyabe widths in CeIrIn$_5$ (squares)
and CeRh$_{0.25}$Ir$_{0.75}$In$_5$ (circles) are temperature
independent across superconducting transition temperatures $T_{\rm c}$.

Fig.2.(a) Corrected Up-Down asymmetries in CeRh$_{0.5}$Ir$_{0.5}$In$_5$
at several temperatures, showing the gradual evolution of 
the spin relaxation function from a static Gaussian
Kubo-Toyabe to oscillating and longitudinal components. The muon spins
in these spectra were rotated to approximately 50$^{\circ}$ from the $c$
axis.
(b) Low temperature Back-Forward asymmetries at $T$=0.1 and 1.15~K
with muon spins oriented along the $c$ axis.  Low temperature
spectra taken in the dilution refridgerator are virtually identical.

Fig.3. The temperature dependence of the frequency $\nu_1$
(circles) and total oscillating amplitude (triangles) in two samples
of CeRh$_{0.5}$Ir$_{0.5}$In$_{5}$.
No change in the relaxation function is seen on crossing into the
superconducting state at $T_{\rm c}$=0.8~K. Filled symbols
correspond to sample 1.

\clearpage
\begin{figure}
\epsfig{figure=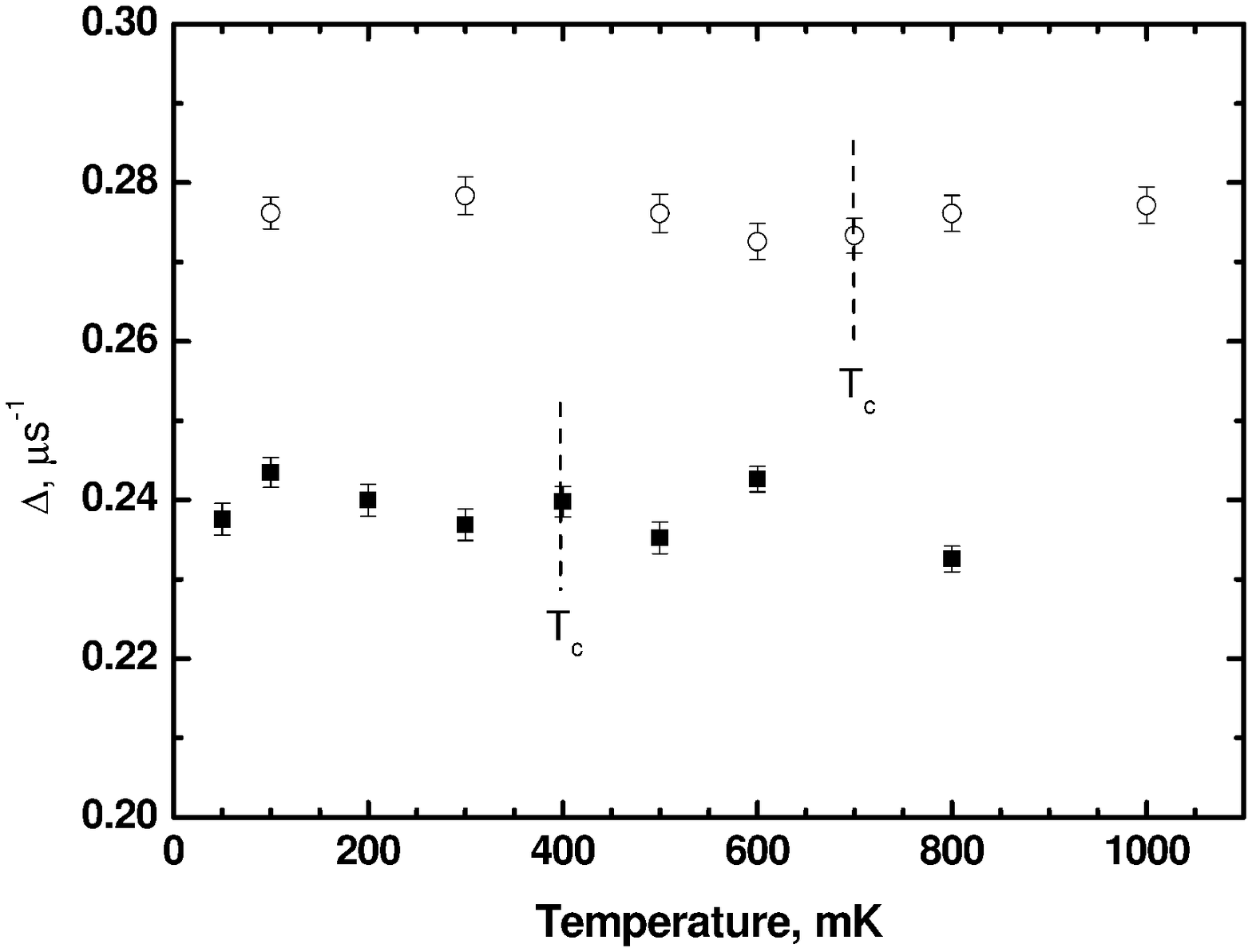}
\label{fig1}
\end{figure}

\clearpage
\begin{figure}
\epsfig{figure=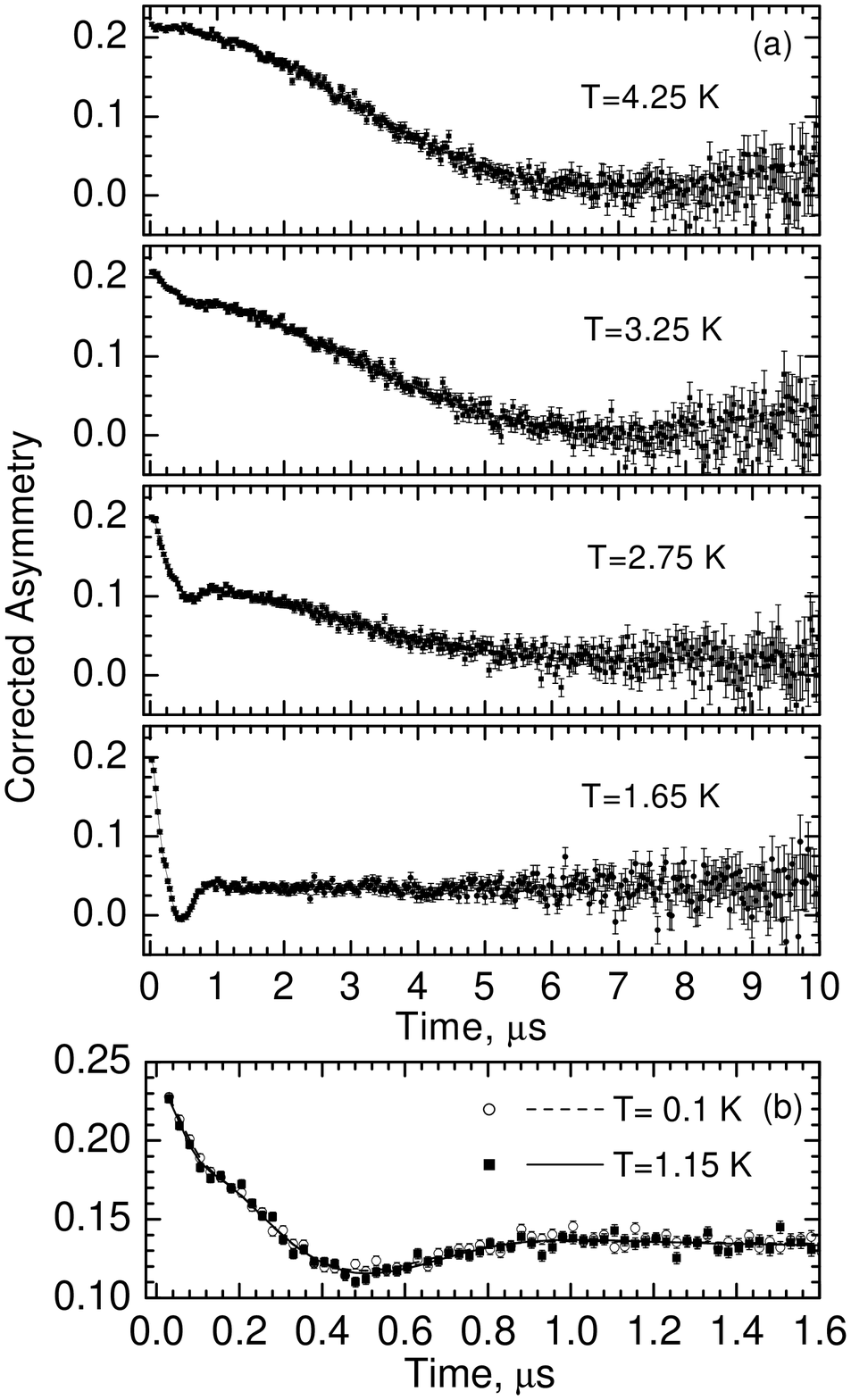}
\label{fig2}
\end{figure}

\clearpage
\begin{figure}
\epsfig{figure=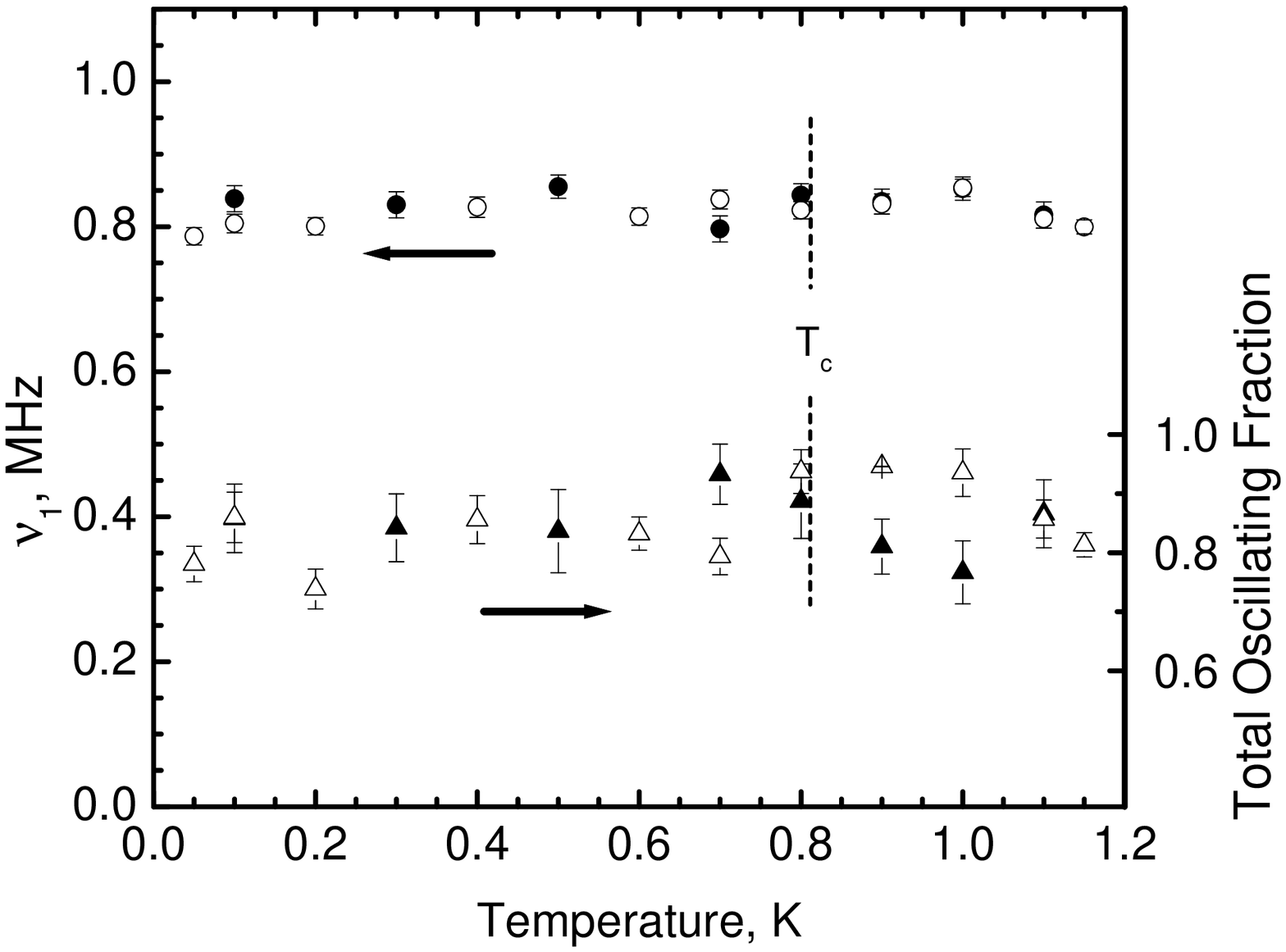}
\label{fig3}
\end{figure}

\end{document}